# COSMIC POLARIZATION ROTATION, COSMOLOGICAL MODELS, AND THE DETECTABILITY OF PRIMORDIAL GRAVITATIONAL WAVES

WEI-TOU NI

*Center for Gravitation and Cosmology, Purple Mountain Observatory, Chinese Academy of Sciences, Nanjing 210008, China   wtni@pmo.ac.cn*
*National Astronomical Observatories, Chinese Academy of Sciences, Beijing 100012, China*

CMB (Cosmic Microwave Background) polarization observations test many aspects of cosmological models. Effective pseudoscalar-photon interaction(s) would induce a rotation of linear polarization of electromagnetic wave propagating with cosmological distance in various cosmological models. CMB polarization observations are superb tests of these models and have the potential to discover new fundamental physics. Pseudoscalar-photon interaction is proportional to the gradient of the pseudoscalar field. From phenomenological point of view, this gradient could be neutrino number asymmetry, other density current, or a constant vector. In these situations, Lorentz invariance or CPT may effectively be violated. In this paper, we review these results and anticipate what more precise observations can tell us about fundamental physics, inflation, etc. Better accuracy in CMB polarization observation is expected from PLANCK mission to be launched this year. Dedicated CMB polarization observers like B-Pol mission, CMBpol mission and LiteBIRD mission would probe this fundamental issue more deeply in the future. With these sensitivities, cosmic polarization rotations from effective pseudoscalar-photon interaction, Faraday polarization rotations from primordial and large-scale magnetic field, and tensor modes effects would have chances to be detected and distinguished. The subtracted tensor-mode effects are likely due to primordial gravitational waves. We discuss the direct detectability of these primordial gravitational waves using space GW detectors.

## 1  Introduction

Since the first successful polarization observation of the cosmological microwave background (CMB) in 2002 by DASI [1] (Degree Angular Scale Interferometer), there have been a number of observations [2-5] with better precision. These observations give the statuses of the Universe at the last Rayleigh scattering together with propagation effects. From the 5-year WMAP data combined with the distance measurements from the Type Ia supernovae (SN) and the Baryon Acoustic Oscillations (BAO) in the distribution of galaxies, the 6 parameters of the ΛCDM model are determined. From these, Komatsu et al. [5] derived that the reionization redshift is $z_{reion} = 10.9 \pm 1.4$. With the WMAP data combined with BAO and SN, the limit on the tensor-to-scalar ratio is $r < 0.22$ (95% CL), and the tilt of the primordial fluctuation power spectrum $n_s > 1$ is disfavored even when gravitational waves are included; this constrains the models of inflation that can produce significant gravitational waves, such as chaotic or power-law inflation models, or a blue spectrum, such as hybrid inflation models [5].

In this paper, we address to two issues. If there is a modification of electromagnetic propagation law due to a phenomenological pseudoscalar-photon interaction, would it be detected via the CMB polarization observation in the next generation detectors? If the tensor mode is detected via the CMB polarization, would primordial gravitational waves generated in the inflation period be directly detectable by space gravitational-wave detectors. In Section 2, we review the phenomenological pseudoscalar-photon interaction and the cosmic polarization rotation. In Section 3, we discuss CMB polarization observation and compile an update of constraints on cosmic polarization rotation to our previous review [6]. In Section 4, we discuss relevant cosmological models. In Section 4, we review briefly the primordial gravitational waves. In Section 5, we discuss the direct detectability of primordial gravitational waves using space detectors.

## 2  Cosmic Polarization Rotation

In the study of the relations among equivalence principles, we have arrived at the following modified electromagnetic Lagrangian density in gravity:

$$L_I = -(1/(16\pi))\, g^{ik}\, g^{jl}\, F_{ij}\, F_{kl}\, (-g)^{1/2} - (1/(16\pi))\, \varphi\, e^{ijkl}\, F_{ij}\, F_{kl} - A_k j^k\, (-g)^{1/2}$$
$$-\Sigma_I m_I (ds_I)/(dt)\, \delta(\mathbf{x}-\mathbf{x}_I), \qquad (1)$$

where $g_{ij}$ is the metric, $\varphi$ a (pseudo)scalar field or a (pseudo)scalar function of other fields, and $j^k$, $F_{ij} \equiv A_{j,i} - A_{i,j}$ have the usual meaning for electromagnetism [7-9]. $e^{ijkl}$ is the completely antisymmetric symbol with $e^{0123} = 1$ [7-9]. Since $\varphi$ is a scalar function, any constant factor in the second term of the interaction Lagrangian $L_I$ could be absorbed.

The second term $L_{I\varphi}$ in the Lagrangian (1)

$$L_{I\varphi} = -(1/(16\pi))\, \varphi\, e^{ijkl}\, F_{ij}\, F_{kl}, \qquad (2)$$

gives a pseudoscalar-photon interaction. Modulo a divergence, (3) is equivalent to [7-9]

$$L_I = (1/(8\pi))\, \varphi_{,i}\, e^{ijkl}\, A_j\, F_{kl}, \quad \text{(mod div)}. \qquad (3)$$

Recent discussions on (3) can be found in Jackiw [10], and Hehl and Obukhov [11]. The special case $\varphi_{,i} = \text{constant} = V_i$ is considered by Carroll, Field and Jackiw [12].

For the theory (1), the electromagnetic wave propagation equation is

$$F^{ik}{}_{,k} + e^{ikml}\, F_{km}\, \varphi_{,l} = 0, \qquad (4)$$

in a local inertial (Lorentz) frame of the $g$-metric. Analyzing the wave into Fourier components, imposing the radiation gauge condition, and solving the dispersion eigenvalue problem, we obtain $k = \omega + (n^\mu \varphi_{,\mu} + \varphi_{,0})$ for right circularly polarized wave and $k = \omega - (n^\mu \varphi_{,\mu} + \varphi_{,0})$ for left circularly polarized wave in the eikonal approximation [7]. Here $n^\mu$ is the unit 3-vector in the propagation direction. The group velocity is

$$v_g = \partial\omega/\partial k = 1, \qquad (5)$$

independent of polarization, hence there is no birefringence. For the right circularly polarized electromagnetic wave, the propagation from a point $P_1 = \{x_{(1)}{}^i\} = \{x_{(1)}{}^0; x_{(1)}{}^\mu\} = \{x_{(1)}{}^0, x_{(1)}{}^1, x_{(1)}{}^2, x_{(1)}{}^3\}$ to another point $P_2 = \{x_{(2)}{}^i\} = \{x_{(2)}{}^0; x_{(2)}{}^\mu\} = \{x_{(2)}{}^0, x_{(2)}{}^1, x_{(2)}{}^2, x_{(2)}{}^3\}$ adds a phase of $\alpha = \varphi(P_2) - \varphi(P_1)$ to the wave; for left circularly polarized light, the added phase will be opposite in sign [7]. Linearly polarized electromagnetic wave is a superposition of circularly polarized waves. Its polarization vector will then rotate by an angle $\alpha$. Locally, the polarization rotation angle can be approximated by

$$\alpha = \varphi(P_2) - \varphi(P_1) = {}_i\Sigma_0{}^3 \, [\varphi_{,i} \times (x_{(2)}{}^i - x_{(1)}{}^i)] = {}_i\Sigma_0{}^3 \, [\varphi_{,i}\Delta x^i] = \varphi_{,0}\Delta x^0 + [{}_\mu\Sigma_1{}^3 \varphi_{,\mu}\Delta x^\mu]$$
$$= {}_i\Sigma_0{}^3 \, [V_i \Delta x^i] = V_0 \Delta x^0 + [{}_\mu\Sigma_1{}^3 V_\mu \Delta x^\mu]. \qquad (6)$$

The rotation angle in (6) consists of 2 parts -- $\varphi_{,0}\Delta x^0$ and $[{}_\mu\Sigma_1{}^3 \varphi_{,\mu}\Delta x^\mu]$. For light in a local inertial frame, $|\Delta x^\mu| = |\Delta x^0|$. In Fig. 2, space part of the rotation angle is shown. The amplitude of the space part depends on the direction of the propagation with the tip of magnitude on upper/lower sphere of diameter $|\Delta x^\mu| \times |\varphi_{,\mu}|$. The time part is equal to $\Delta x^0 \, \varphi_{,0}$. ($\nabla \varphi \equiv [\varphi_{,\mu}]$) When we integrate along light (wave) trajectory in a global situation, the total polarization rotation (relative to no $\varphi$-interaction) is again $\Delta\varphi = \varphi_2 - \varphi_1$ for $\varphi$ is a scalar field where $\varphi_1$ and $\varphi_2$ are the values of the scalar field at the beginning and end of the wave. When the propagation distance is over a large part of our observed universe, we call this phenomenon cosmic polarization rotation [6].

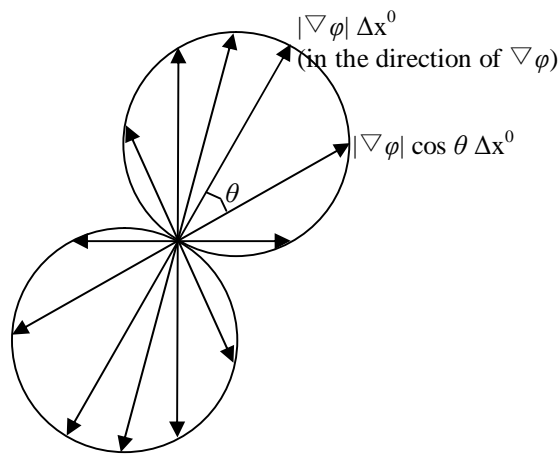

Figure 1. Space contribution to the local polarization rotation angle -- $[{}_\mu\Sigma_1{}^3 \varphi_{,\mu}\Delta x^\mu] = |\nabla\varphi| \cos\theta \, \Delta x^0$. The time contribution is $\varphi_{,0} \Delta x^0$. The total contribution is $(|\nabla\varphi| \cos\theta + \varphi_{,0}) \Delta x^0$. ($\Delta x^0 > 0$).

In the CMB polarization observations, there are variations and fluctuations. The variations and fluctuations due to scalar-modified propagation can be expressed as $\delta\varphi(2) - \delta\varphi(1)$, where 1 denotes a point at the last scattering surface in the decoupling epoch and 2 observation point. $\delta\varphi(2)$ is the variation/fluctuation at the last scattering surface. $\delta\varphi(1)$ at the present observation point is zero or fixed. Therefore the covariance of fluctuation $<[\delta\varphi(2) - \delta\varphi(1)]^2>$ gives the covariance of $\delta\varphi^2(2)$ at the last scattering surface. Since our Universe is isotropic to $\sim 10^{-5}$, this covariance is $\sim (\xi \times 10^{-5})^2$ where the parameter $\xi$ depends on various cosmological models.

## 3  CMB Polarization Observation and Constraints on Cosmic Polarization Rotation

In this section, we review and compile the constraints on the cosmic polarization rotation angle from various analyses of CMB polarization observations. Constraints prior to the CMB observations are discussed in [6].

In 2002, DASI microwave interferometer observed the polarization of the cosmic background [1]. E-mode polarization is detected with 4.9 σ. The TE correlation of the temperature and E-mode polarization is detected at 95% confidence. This correlation is expected from the Raleigh scattering of radiation. However, with the (pseudo)scalar-photon interaction (2)/(3), the polarization anisotropy is shifted differently in different directions relative to the temperature anisotropy due to propagation; the correlation will then be downgraded. In 2003, from the first-year data (WMAP1), WMAP found that the polarization and temperature are correlated to more than 10 σ [2]. This gives a constraint of about $10^{-1}$ for $\Delta\varphi$ [13, 14].

Further results [3-5] and analyses [5, 15-22] of CMB polarization observations came out after 2006. In Table 1, we update our previous compilation of [6]. Although these results look different at 1 σ level, they are all consistent with null detection and with one another at 2 σ level. We turn to the interpretation of cosmic polarization rotation in the next section.

Both magnetic field and potential new physics affect the propagation of CMB propagation and generate BB power spectra from EE spectra of CMB. The Faraday rotation due to magnetic field is wavelength dependent while the cosmic polarization rotation due to effective pseudoscalar-photon interaction is wavelength-independent. This property can be used to separate the two effects. With the tensor mode generated by these two effects measured and subtracted, the remaining tensor mode perturbations are likely due to primordial (inflationary) gravitational waves (GWs). We discuss the direct detectability of these primordial GWs using space GW detectors in Section 6.

Table 1. Constraints on cosmic polarization rotation from CMB (cosmic microwave background).

| Reference | Constraint [mrad] | Source data |
|---|---|---|
| Ni [13, 14] | ±100 | WMAP1 [2] |
| Feng, Li, Xia, Chen, and Zhang [15] | −105 ± 70 | WMAP3 [3] & BOOMERANG (B03) [4] |
| Liu, Lee, Ng [16] | ±24 | BOOMERANG (B03) [4] |
| Kostelecky and Mews [17] | 209 ± 122 | BOOMERANG (B03) [4] |
| Cabella, Natoli and Silk [18] | −43 ± 52 | WMAP3 [3] |
| Xia, Li, Wang, and Zhang [19] | −108 ± 67 | WMAP3 [3] & BOOMERANG (B03) [4] |
| Komatsu, *et al.* [5] | −30 ± 37 | WMAP5 [5] |
| Xia, Li, Zhao, and Zhang [20] | −45 ± 33 | WMAP5 [5] & BOOMERANG (B03) [4] |
| Kostelecky and Mews [21] | 40 ± 94 | WMAP5 [5] |
| Kahniashvili, Durrer, and Maravin [22] | ± 44 | WMAP5 [5] |

## 4  Cosmological Models

Feng *et al.* [15] proposed CPT violation and dynamical dark energy. In a more recent paper, Li *et al.* [23] considered baryo/leptogenesis with cosmological CPT violation as a possible cause and gave a 1σ limit on their fermion current-curvature coupling parameter δ = − 0.011 ± 0.007. With the results of Xia *et al.* [20], coupling parameter δ would be decreased by a factor about 2. Liu, Lee and Ng gave constraints on the coupling between the quintessence and the pseudoscalar of electromagnetism [16]. Kostelecky and Mews extended their SME [17] (Standard Model Extension) whose electromagnetic sector is the same as that in (1) of [9] with the gravitational constitutive tensor set to constant, to include some higher order terms, and gave constraints on various terms from BOOMERANG. In [21], they did their analysis using WMAP5 data [5]. The most precise constraint is on one of the SME parameter which gives cosmic polarization rotation $\Delta\varphi$. Their constraints [17, 21] are shown in Table I.

Geng, Ho and Ng proposed a new type of effective interactions in terms of the CPT-even dimension-six Chern-Simons-like term to generate the cosmic polarization rotation, and used the neutrino number asymmetry to induce a non-zero polarization rotation angle in the data of the CMB polarization [24]. They found that the rotation effect can be of the order of magnitude of 10-100 mrad or smaller.

## 5  Primordial GWs

The intensity of a stochastic primordial background of GWs is usually characterized by the dimensionless quantity

$$\Omega_{GW}(f) = (1/\rho_c) \, (d\rho_{GW}/d\log f), \tag{7}$$

with $\rho_{GW}$ the energy density of the stochastic GW background and $\rho_c$ the present value of the critical density for closing the universe in general relativity. Primordial GWs are believed to be generated in the early Universe. Their theoretical predictions have a large range. However, we can use experiments to give upper bounds. Figure 2 shows these upper bounds together with sensitivity of various GW detectors based on [25, 26]. Most notably is the CMB bound at lower frequency. The tilt of the primordial fluctuation power spectrum $n_s$ is determined to be $n_s = 0.960$ (+0.014−0.013). The primordial GWs are believed to have a spectrum close to this spectrum and n = 1 serves a good bound in the figure. Other bounds are explained in [25]. Optimistic predictions of primordial GWs' $\Omega_{GW}$ are in the range of $10^{-15}$-$10^{-18}$.

## 6  The Detectability of Primordial GWs

For detection of primordial gravitational waves in space, one may go to frequencies lower or higher than the LISA bandwidth where there are potentially less foreground astrophysical sources [27] to mask detection. DECIGO and Big Bang Observer look for gravitational waves in the higher range while ASTROD and Super-ASTROD [26] looks for gravitational waves in the lower range.

As armlength *L* increases, the sensitivity curve is shifted downward. The minimum detectable intensity of a stochastic GW background $\Omega_{GW}^{min}(f)$ is proportional to detector noise power spectral density $S_n(f)$ times frequency to the third power [25]. That is

$$h_0^2 \, \Omega_{GW}^{min}(f) \sim \text{const.} \times f^3 \, S_n(f), \tag{8}$$

where $h_0$ is the present Hubble constant $H_0$ divided by 100 km s$^{-1}$ Mpc$^{-1}$. Hence, with the same strain sensitivity, lower frequency detectors have an $f^{-3}$-advantage over the higher frequency detectors. This is the main reason why ASTROD and Super-ASTROD can probe deep into inflation GW region. In figure 2, compared to LISA, ASTROD has 27,000 times ($30^3$) better sensitivity due to this reason. For Super-ASTROD there is an

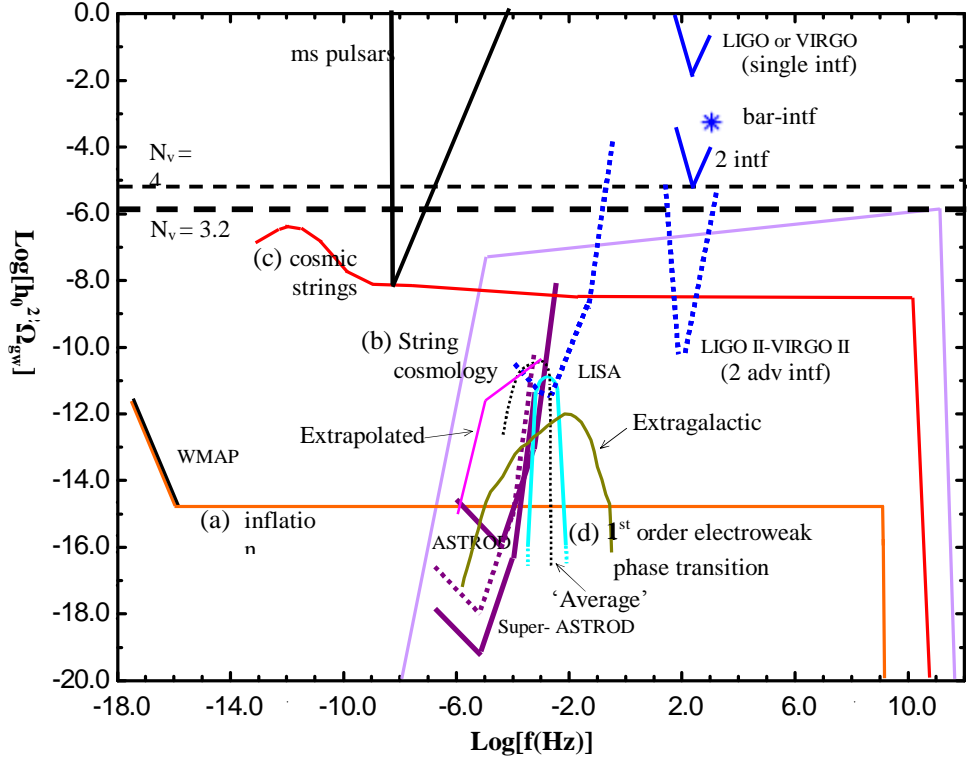

Figure 2. The stochastic backgrounds optimistically predicted with bounds and GW detector sensitivities. (Adapted from Figure 3 and Figure 4 of [25] with sensitivity curves of ASTROD and Super-ASTROD added; the extragalactic background and the extrapolated background curves are from [26]; see [25-27] and Section 6 for explanations).

additional 125 ($5^3$) times gain in sensitivity (the dotted line in between the ASTROD and Super-ASTROD sensitivity curves) compared to ASTROD due to longer armlength if the laser power and $S_n(f)$ stays the same. However, for baseline Super-ASTROD, the laser power is increased by 15 fold, and hence, the shot noise floor for strain sensitivity is lowered by a factor of 3.87. In terms of noise power spectral density, it is lowered by a factor of 15. Hence, the total gain in sensitivity due to $f^3 S_n(f)$ is a factor of 1875. In the study of GW background from cosmological compact binaries, Farmer and Phinney [27] showed that this background $\Omega_{gw}^{cb}(f)$ rolls off in the 1-100 µHz frequency region from $10^{-13}$ at 100 µHz to $10^{-17}$ level at 1 µHz. Therefore, Super-ASTROD will be able to detect this background and there is still ample room for detecting the primordial/inflationary GWs with optimistic amplitudes. For ASTROD, although the sensitivity curve extends well below the optimistic inflation line, the background from binaries [27] in the ASTROD bandwidth is as large as the most optimistic inflation signals. Signal separation methods should be pursued. If relic primordial/inflationary

GW has a different spectrum than that cosmological compact binaries, their signals can be separated to certain degree. With this separation, Super-ASTROD will be able to probe deeper into the stochastic waves generated during inflation with less optimistic parameters. More theoretical investigations in this respect are needed.

## 7  Discussion and Outlook

Better accuracy in CMB polarization observation is expected from PLANCK mission to be launched this year. Dedicated CMB polarization observers like B-Pol mission, CMBpol mission and LiteBIRD mission would improve the sensitivity further. Direct detection of Primordial GWs may be possible with Big Bang Observer, DECIGO, ASTROD and Super-ASTROD. These development would probe the fundamental issues discussed in this paper more deeply in the future.

We would like to thank the National Natural Science Foundation of China (Grant Nos. 10778710 and 10875171) and the Foundation of Minor Planets of Purple Mountain Observatory for support.